\documentclass[final,3p,times,twocolumn]{elsarticle}

\usepackage{amssymb,amsmath}
\usepackage{bm}

\usepackage{booktabs}
\usepackage{braket}
\usepackage{url}
\usepackage[english]{babel}
\usepackage{graphics}
\usepackage{tabularx}
\usepackage{rotating}
\usepackage{float}

\graphicspath{{./figures/}}

\journal{Physics Letters A}

\begin{document}

\begin{frontmatter}

\title{Efimov effect in non-integer dimensions induced by an external field}

\author[1]{E. Garrido}
\ead{e.garrido@csic.es}
\author[2]{A.S. Jensen}

\address[1]{Instituto de Estructura de la Materia, IEM-CSIC, Serrano 123, E-28006 Madrid, Spain}
\address[2]{Department of Physics and Astronomy, Aarhus University, DK-8000 Aarhus C, Denmark}

\date{\today}

\begin{abstract}
The Efimov effect can be induced by means of an external
deformed one-body field that effectively reduces the allowed spatial
dimensions to less than three.  To understand this new mechanism,
conceptually and practically, we employ a formulation using
non-integer dimension, which is equivalent to the strength of an external oscillator field.
The effect most clearly appears when the crucial two-body systems are unbound in three, 
but bound in two, dimensions.  We discuss energy variation, conditions for occurrence,
and number of Efimov states, as functions of the dimension.  We use practical
examples from cold atom physics of $^{133}$Cs-$^{133}$Cs-$^{133}$Cs,
$^{87}$Rb-$^{87}$Rb-$^{87}$Rb, $^{133}$Cs-$^{133}$Cs-$^{6}$Li, and
$^{87}$Rb-$^{87}$Rb-$^{39}$K.  Laboratory tests of the effect can be
performed with two independent parameters, i.e. the external one-body
field and the Feshbach two-body tuning.  The scaling and (dis)appearance of 
these Efimov states occur precisely as already found in three dimensions.
\end{abstract}


\begin{keyword}
Efimov effect \sep confinement of quantum systems \sep $d$-dimensional calculations

\end{keyword}
\end{frontmatter}

\section{Introduction}                            

The Efimov effect was suggested theoretically about fifty years ago
for three-body systems \cite{efi70}.  For the occurrence, at least
two of the three pair-interactions must have nearly zero energy.
Occurrence is optimized by having identical constituents, where only
one two-body interaction is involved.  However, although in principle
possible, to find in nature two particles, identical or not, bound by 
zero energy would be very rare.

In cold atom physics this problem was overcome about 25 years ago
with an original method, that is the technique of controlled tuning of 
the effective two-body interaction by coupling via a Feshbach
resonance \cite{koh06,blo08,chi10,den16}.  Subsequently, properties
and consequences of the Efimov effect have been extensively studied
theoretically \cite{nie01,jen04,fed03,bra06,gar18,mik15} and also
early established experimentally
\cite{kra06,zac09,pol09,gro09,huc09,ott08,wen09}.

An important fact concerning the Efimov effect is that, whereas 
present in three dimensions (3D), is absent in two (2D)
\cite{bru79,lim80,nie97,vol13,nai17}.  The reason is that in 3D a 
finite attraction is required for binding a two-body system, but
in 2D an infinitesimal attraction is sufficient \cite{sim76,vol11}.
These points emphasize the variation between
dimensions, and in fact triggered previous investigations of 
dimensional transitions by different methods \cite{lev14,yam15,san18,ros18}.
In \cite{nis08} it is shown how a two-species Fermi gas where one species is confined 
in a two- or one-dimensional space, while the other one is free in 3D, 
can lead, depending on the mass ratio, to systems showing the Efimov
effect. This is the confinement-induced Efimov effect introduced in \cite{nis09},
although no details are given on the influence from continuous external squeezing.
A short review of the effects of the confinement in mixed dimensions is
given in \cite{nai17}. 

A novel three-body method applicable to non-integer dimensions, $d$,
was presented about two decades
ago \cite{nie01}. One spectacular prediction was that the Efimov
effect is only possible in dimensions between $2.3$ and $3.8$.  This
$d$-method was recently implemented to investigate the confinement of
two- and three-body systems \cite{gar19a,chr18,gar19b,gar20}.
Practical calculations for non-integer dimensions are
precisely as easy, or difficult, as in the usual 3D-space.  The necessary
relation to ordinary physics of integer-based dimensions is
also available \cite{gar19a,gar19b}.  The  translation involves an 
external deformed field \cite{moe19},
which is shown to be  equivalent to the non-integer
dimensional treatment \cite{gar20}. 

A detailed comparison of the actual 
wave function obtained introducing explicitly the external
field and the deformed wave function extracted from the $d$-method
is also given in \cite{gar20} for different two-body potentials. It is 
shown how both wave functions are very much equivalent for
both, well-bound and weakly bound, three-body systems. Thus, we can 
employ the simple method and interpret in terms of a deformed external one-body field.

The purpose of this letter is to demonstrate the power of the 
$d$-method by exhibiting a spectacular consequence of applying an
external manageable \cite{arm15} oscillator field to a three-body
system.  Assume the two-body attraction in 3D is too weak to bind. We then
employ a squeezing one-body oscillator field on the $z$-direction,
while leaving the $x$ and $y$-coordinates untouched. The two-body
systems are then forced to move towards 2D, where they at some point become
bound through the unchanged two-body attractions.  At this point of
zero binding, the Efimov effect appears, which is at a non-integer
dimension, or equivalently, at a certain strength of the external field.

Of course, the Feshbach resonance technique is still available, and 
the combination with an independent tuning of the deformed external field might prove
convenient.  The two independent parameters can be used
to approach the optimal situation with three different subsystems at
zero energy.  This is not possible with only one parameter, neither
Feshbach nor external field. 

 In this letter, we first describe the $d$-method and give enough details 
to allow calculations and estimates.
We then present results for use on specific cold atom gases, where
the Efimov effect can be manipulated to appear for non-integer
dimensions.

\section{The $d$-method} 

In Efimov's original formulation the key to understanding is that a
long-distance 
effective attractive potential with an inverse-distance-square dependence
arises for three particles with a strength more negative
than a given critical value.  The crucial potential term has
the same form as a centrifugal barrier, which in ordinary
three-dimensional space is positive.
How and when this occurs is  well established by some pathological
conditions obeyed by the three constituent particles and their
interactions.  Although these occurrence conditions are special, they
have been simulated in laboratories and a number of derived
consequences experimentally tested.  

To describe and understand in details it is convenient to adopt the
formalism of the hyperspherical adiabatic expansion method, where the
hyperradius, $\rho$, becomes the crucial length coordinate entering in the 
corresponding differential Schr\"{o}dinger equation.  The essential
result, derived in many previous publications, e.g. \cite{nie01}, takes
the following form for each adiabatic channel:
\begin{equation}
\left( 
 -\frac{\partial^2}{\partial \rho^2}+\frac{\lambda^{(d)}(\rho)+(d-1)^2-\frac{1}{4}}{\rho^2}
- \frac{2mE_{3b}^{(d)}}{\hbar^2}
\right) f^{(d)}(\rho) = 0,
\label{rads1}
\end{equation}
where $d$ is the dimension parameter ($2 \le d \le 3$),
$E_{3b}^{(d)}$ is the three-body energy in $d$ dimensions, $m$ is a normalization mass
disappearing in all observable quantities,  and $\lambda^{(d)}$ is the
$\rho$-dependent eigenvalue from the angular Schr\"{o}dinger (or Faddeev) equation. 
The reduced, $f^{(d)}$, and total, $F^{(d)}$, radial wave functions
are related by
\begin{eqnarray} \label{db20}
 f^{(d)}(\rho) =  \rho^{d-1/2} F^{(d)}(\rho) \; .
\end{eqnarray}

Only the diagonal terms are included in Eq.(\ref{rads1}), since the
couplings between the adiabatic channels are unimportant as they
vanish at the decisive large distances.  Here we should remember that
all such formula refer to one decoupled adiabatic channel. 

The Efimov effect occurs  for each channel for the particular dimension $d=d_E$ for
which the numerator of the effective potential in Eq.(\ref{rads1})  is, in the 
large-$\rho$ limit, constant and less than $-1/4$.  From Eq.(\ref{rads1}) 
we  see that this condition means that  $\xi_{d_E}^2 < 0$, or $\xi_{d_E}=i|\xi_{d_E}|$, where we have defined
\begin{eqnarray} \label{ksi}
 \xi_{d_E}^2 = \lambda^{(d_E)}_\infty+(d_E-1)^2,
\label{xi}
\end{eqnarray}
with  $\lambda_\infty^{(d_E)}=\lambda^{(d_E)}(\rho=\infty)$.

The bound state solutions to Eq.(\ref{rads1}) for constant
$\lambda^{(d_E)}=\lambda^{(d_E)}_\infty$ are $f^{(d_E)}(\rho) \propto \sqrt{\kappa_{d_E} \rho}
K_{i|\xi_{d_E}|}(\kappa_{d_E}\rho)$,  with $\kappa_{d_E} = \sqrt{-2mE_{3b}^{(d_E)}/\hbar^2}$.
The modified Bessel function of second kind, $K_{i|\xi_{d_E}|}$,
decreases exponentially at large distances as it should for a bound
state.  For small distances we have
instead $K_{i|\xi_{d_E}|} \propto \sin(|\xi_{d_E}| \ln(\kappa_{d_E}\rho))$.

It is important to keep in mind that in actual calculations the $\lambda^{(d_E)}$-function 
is constant only  over a $\rho$-interval limited by a scattering length, $|a_{d_E}^{av}|$, in practice
finite, and defined as an average of
the three $d$-dimensional two-body scattering lengths involved in the three-body system \cite{fed03}. 
The condition $\xi_{d_E}^2<0$, or $ \lambda^{(d_E)}_\infty<-(d_E-1)^2$, 
and therefore the Efimov effect, requires that  at least two of the two-body scattering
lengths are numerically very large, which is equivalent to having close to zero energy 
in at least two of the  two-body subsystems. For $\rho$ larger than $|a_{d_E}^{av}|$ the
$\lambda$-functions can no longer support bound states.  The number, $N_E$, of Efimov states 
that can be held by a large, but finite, scattering length $a_{d_E}^{av}$,
can be estimated by counting the number of nodes in $K_{i|\xi_{d_E}|}$ between the ground 
state size, $\rho_0$, and $|a_{d_E}^{av}|$, that is \cite{nie01}
\begin{equation}
 N_E \approx \frac{|\xi_{d_E}|}{\pi} \ln\big(\frac{|a_{d_E}^{av}|}{\rho_0}\big) \;.
\label{num}
\end{equation}

The usual Efimov scaling  for energies and root-mean-square (rms) radii still applies, which is \cite{nie01}
\begin{equation}
 \frac{E_n}{E_{n+1}} = \frac{\langle\rho^2\rangle_{n+1}}{\langle\rho^2\rangle_{n}} = e^{2\pi/|\xi_{d_E}|} \;,
\label{scal}
\end{equation}
where $n$ labels the different states in the Efimov series.

\section{External field translation} 

The formulation in terms of the exceedingly intuitive dimension
parameter, $d$, is very efficient and convenient for theoretical
calculations.  However, a relation to laboratory controlled
observable variables is needed.  This has recently become available in
investigations where the particles are confined, say in the
$z$-coordinate, by an external one-body oscillator field.

These calculations, where the deformed squeezing external 
oscillator potential with frequency $\omega_{ho}$  is included explicitly, 
are performed in a three-dimensional space, and 
they are related to the above $d$-results as shown in Ref.\cite{gar20}, 
which is:
\begin{equation}
 \frac{\omega_{pp}}{\omega_{ho}} = \frac{2(d-2)}{(d-1)(3-d)}  \;,
\label{bdrel}
\end{equation}
where $\omega_{pp}$ is the frequency of the equivalent two-body oscillator
interaction that, when used in the three-body calculation, gives rise to the same 
rms radius $r_{2D}=\sqrt{2\hbar/(M \omega_{pp})}$,
where $M$ is the total mass, as the original potentials.
In other words, in actual calculations with arbitrary two-body potentials, 
the computed value of $r_{2D}$ permits to obtain $\omega_{pp}$ to be
used in Eq.(\ref{bdrel}). 

It was already shown in \cite{nie01} that for $s$-waves and three
identical bosons the appearance of the Efimov effect requires 
$d>2.298$. This limit translates into $\omega_{ho} < 1.02  \omega_{pp}$, which 
means that a squeezing frequency larger than $1.02  \omega_{pp}$
cannot produce an Efimov effect for three identical bosons.
For one light and two
identical heavy particles the limit value can move up or down depending on 
the mass ratio, as shown in \cite{chr18}. 

After a $d$-calculation, the equivalent
deformed three-dimensional wave function is constructed by scaling  
down the $z$-direction by a factor, $s$, such that
\begin{equation} \label{rhos}
\rho^2 \rightarrow \tilde{\rho}^2 = \rho^2_{\perp} +  \rho^2_{z}/s^2 \; .
\end{equation}
The constant scaling, $s$, is approximately given by \cite{gar20}
\begin{equation}
 \frac{1}{s^2}=\sqrt{1+\frac{\omega_{ho}^2}{\omega_{pp}^2}}=
 \sqrt{1+\left( \frac{(d-1)(3-d)}{2(d-2)}  \right)^2} \;,
\label{estim}
\end{equation}
which connects the dimension $d$ and the deformation, determined by the
scale parameter $s$, of the actual three-dimensional wave function.
We then replace $f^{(d)}(\rho)$ by $f^{(d)}(\tilde{\rho})$, which, after the 
necessary renormalization, allows computations of any desired observable
investigated in the laboratory from knowledge entirely from
$d$-calculations, see \cite{gar19b,gar20} for details.

\section{Interactions and properties}

Realistic numerical investigations need finite-range two-body
potentials.  We have chosen a potential such that the two-body systems 
are unbound in 3D, which will later permit to highlight the emergence of 
Efimov states for $2\leq d \leq 3$. For our purpose any potential
fulfilling these conditions could be used. The conclusions are independent
of the specific potential shape.

For simplicity we choose  a Gaussian radial
shape, $S\exp(-r^2/b^2)$.  This potential shape has been 
frequently used to investigate the Efimov effect, from the
early days of three-body calculations \cite{fed94}, to much
more recent works \cite{kie13,hap19}. Taking  $\hbar^2/(\mu b^2)$ as
energy unit, the properties of the two-body systems are mass-independent.  In 2D the
two-body systems have only one bound state, whose energy,
$E_{2D}^{(2bd)}$, is given in the second row of Table~\ref{tab1}. The
following two rows give $a_{2D}$ and $a_{3D}$, in units of $b$, which are the
scattering lengths, $a_{d}$, for $d=2$ and $d=3$, respectively.

\begin{table}[t]
\begin{tabular}{|c|ccc|} \hline
$m_H/m_L$ &  1/1  (3 ident.)&   133/6   &  87/39  \\ \hline
$E_{2D}^{(2bd)}$  &  $-0.0987$  &  $-0.0987$  & $-0.0987$  \\
$a_{2D}$  &  2.842  &   2.842  &   2.842  \\
$a_{3D}$  & $-2.699$ & $-2.699$  & $-2.699$  \\ \hline
$E_{2D}^{(3gr)}$ &  $-0.508$ &  $-0.341$ & $-0.236$  \\
$E_{2D}^{(3*)}$  &  $-0.101$  &   $-0.167$   &    \\
$E_{2D}^{(3**)}$  &        &   $-0.110$   &    \\
$r_{2D}^{(3gr)}$  &  0.744 &  0.466  &  1.016 \\
$r_{2D}^{(3*)}$  &   6.040  &  1.144   &  \\
$r_{2D}^{(3**)}$  &       &  2.545   &  \\  \hline
\end{tabular}
\caption{We consider a three-body system made of two heavy identical
particles with mass $m_H$ and a light particle with mass $m_L$. The
heavy-heavy interaction is zero. The heavy-light Gaussian
interaction is with range, $b$, and strength $S= - 0.943 \hbar^2/(\mu b^2)$,
where $\mu$ is the reduced mass of the system. For $m_H/m_L=1$ the three
particles and the corresponding interactions are identical. For $m_H/m_L$=1,
133/6, and 87/39 we give two-body ground state energies, $E_{2D}^{(2bd)}$,
and scattering lengths, $a_{2D}$, and $a_{3D}$ in 2D and 3D. The lower part of
the table shows, in 2D, the three-body energies and rms radii of the ground state 
and the existent excited states. Lengths are in units of $b$, and energies in
units of $ \hbar^2/(\mu b^2)$. }
\label{tab1}
\end{table}

We consider three-body systems made of two identical heavy
particles with mass $m_H$ and a light particle with mass $m_L$. The 
interaction between the two heavy particles is put equal to zero,
except for the mass ratio $m_H/m_L=1$, where the three particles are
considered identical. As shown later, the asymmetric  $m_H/m_L=1$
case (one of the interactions equal to zero) is not particularly convenient
for our purpose.
The potential does not by choice bind the three-body systems in 3D,
reflected by the moderately negative scattering length $a_{3D}$. In
contrast, in the three cases shown in Table~\ref{tab1}, the three-body
system has one well bound ground state in 2D. Furthermore, also in 2D,
the mass ratios $m_L/m_H=1$ and $m_H/m_L=133/6$ have one and two bound
excited three-body states, respectively, see \cite{bel12}. The
corresponding energies are given in the lower part of Table~\ref{tab1}
together with the root-mean-square radii for each of the states.

\section{Energy variation}

\begin{figure}[t]
\centering
\includegraphics[width=0.85\linewidth]{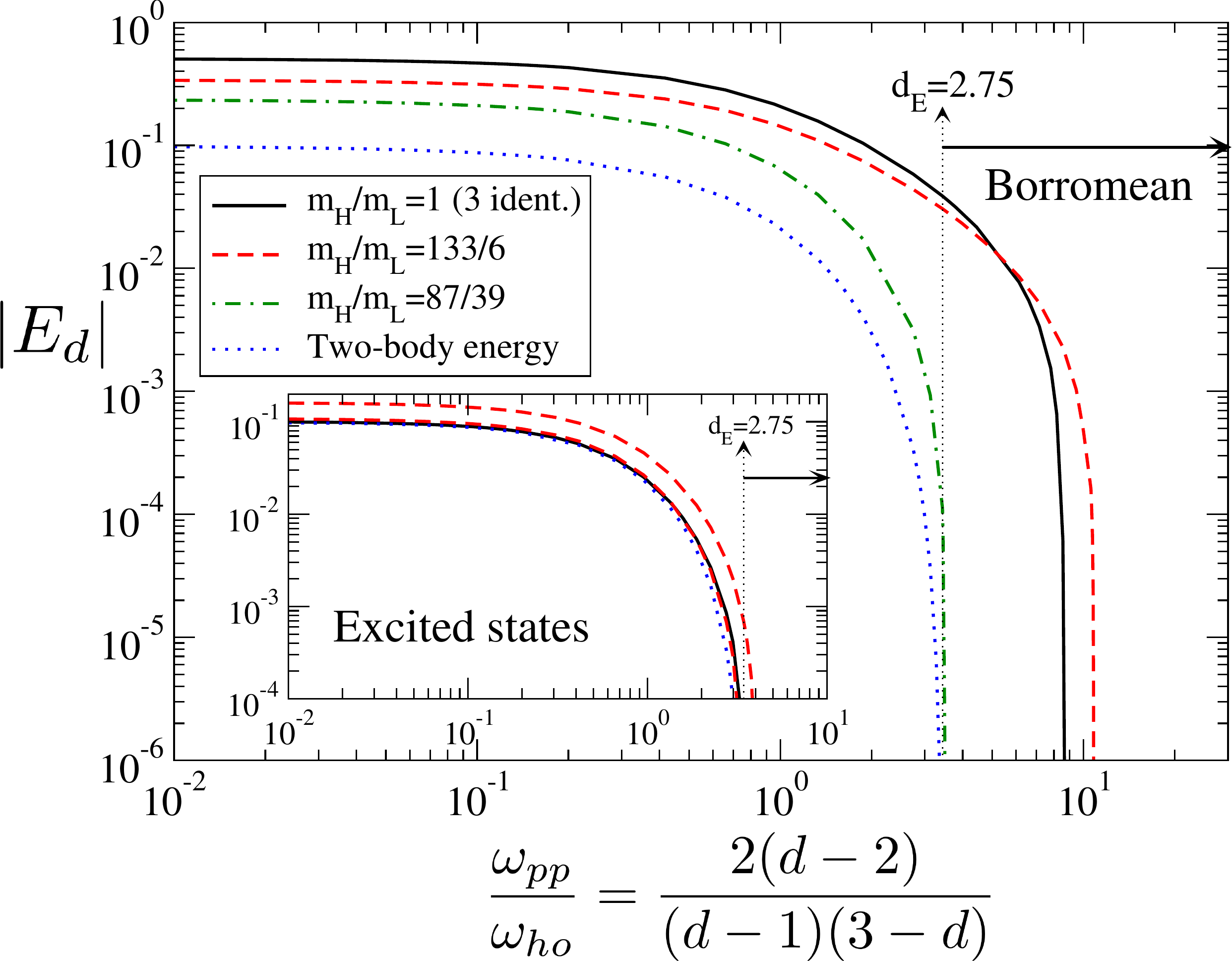} 
\caption{For the three-body states given in the lower
part of Table~\ref{tab1}, absolute value of the bound 
three-body energies,  $|E_d|$ in units of $\hbar^2/(\mu b^2)$, 
as a function of the $d$-function in Eq.(\ref{bdrel}).
The dotted blue curves is  the absolute energy of the 
bound two-body state. The outer and inner panels show,
respectively,  the evolution of the ground state and the 
excited states. The black arrow indicates the region where
the bound three-body states are borromean systems ($d>d_E$). }
\label{fig1}
\end{figure}

Starting from $d=3$, when the dimension is decreased, i.e., when the 
system is progressively squeezed, the bound three-body states for
each system appear from the continuum as shown in Fig.~\ref{fig1},
where we show the binding energies as functions of the $d$-function 
in Eq.(\ref{bdrel}). The outer and inner panels show, respectively,
the evolution of the ground and excited state energies for the
different mass ratios. In both panels the dotted curve gives the
evolution of the heavy-light two-body energy.  The vertical dotted
line at $d_E=2.75$ marks the dimension below which the two-body system
is bound.  Therefore, all the three-body bound states located to the
right of this vertical line have borromean character.

The overall behavior of the energy curves in Fig.~\ref{fig1} is
similar for all the cases.  When confining from 3D to 2D, moving from
right to left in the figure, different bound states appear, that is
the ground states for $d=2.89$, 2.91, and 2.76, for
$m_H/m_L=1$, $133/6$, and $87/39$, respectively.  From this $d$-value,
where the systems are borromean, an increase of the confinement
(decrease of $d$) gives rise to a fast increase of the energy, which
stabilizes following the same trend as the two-body energy.  The
behavior is similar for the excited states, shown in the inner panel,
although the three-body energies are much closer to the two-body ones.
In particular, the energy of the first excited state  for $m_H/m_L=1$
and the second excited state for $m_H/m_L=133/6$, appearing both
at $d$=$2.76$, follow very closely the two-body
energy. The first excited state for $m_H/m_L=133/6$
appears at $d=2.78$.

The dimension $d_E=2.75$, corresponding to zero two-body energy, is
the Efimov point, where infinitely many three-body bound states also
emerge, and soon after disappear again.  However, this $d$-value is
rather arbitrary and the three coinciding Efimov points are only due
to the choice of interactions.  Increasing the two-body attraction, or
equivalently $|a_{3D}|$, would move the corresponding Efimov point to
the right on Fig~\ref{fig1}.  Eventually the well known
Efimov condition would be reached in 3D with infinitely many states
corresponding to zero two-body binding energy at $d=d_E=3$.  By
weakening the attractive two-body potentials, the curves would move to
the left towards 2D.

\begin{figure}[t]
\centering
\includegraphics[width=0.85\linewidth]{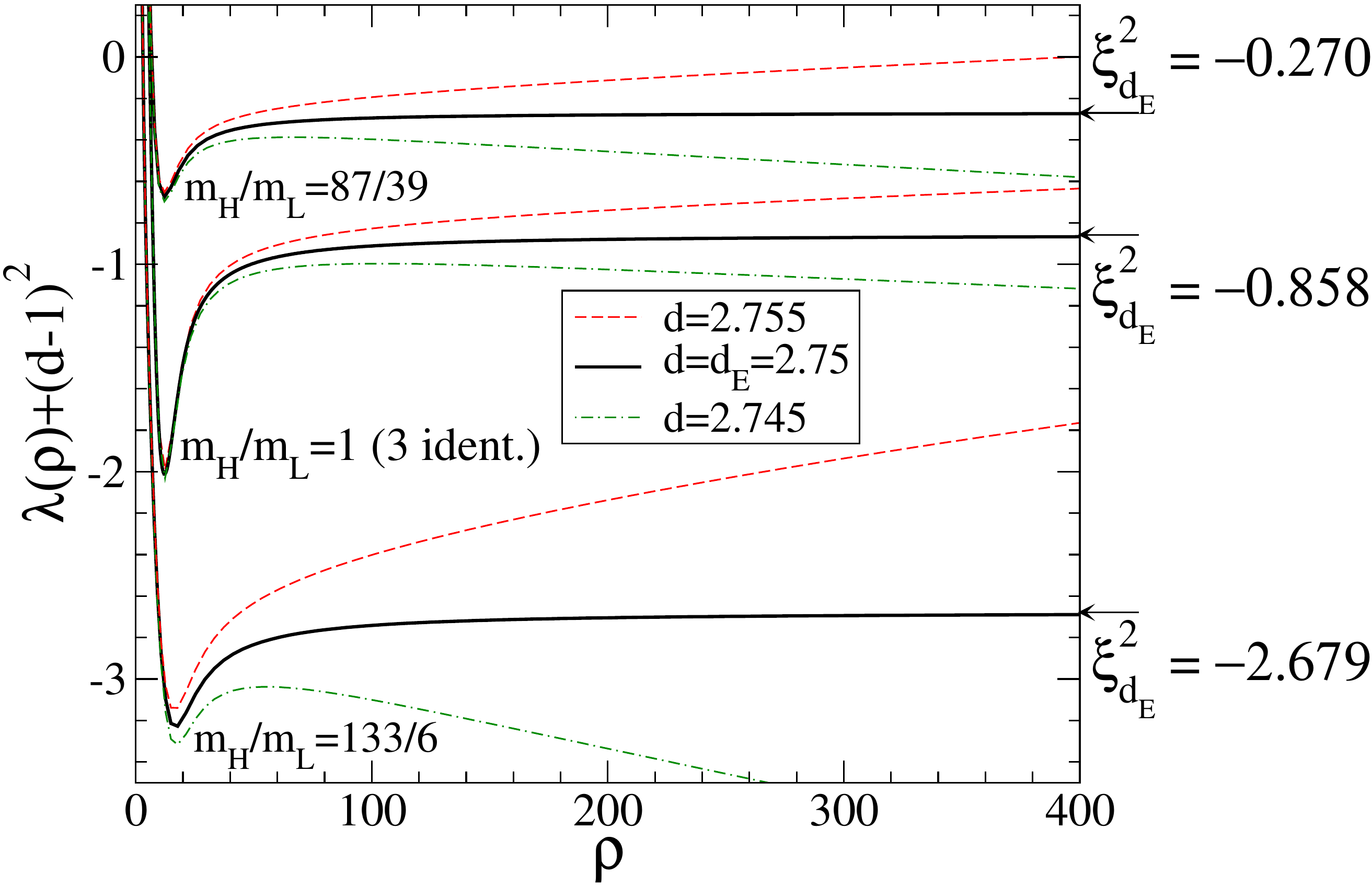}
\caption{The functions $\lambda_n^{(d)}+(d-1)^2$ for the lowest diagonal
potentials as functions of $\rho$ for $d=d_E=2.75$ (solid), 
d=2.755 (dashed) and $d=2.745$ (dot-dashed), for 
mass ratios $m_H/m_L=1/1$ (3 identical particles), $133/6$, and $87/39$.  The
arrows indicate the asymptotic value of $\xi_{d_E}^2$ for $d=d_E$.}
\label{fig2}
\end{figure}

\section{Asymptotic potentials}

To understand the mechanism, we look at the effective potentials in
Eq.(\ref{rads1}). In Fig.~\ref{fig2} the solid curves show
$\lambda^{(d)}(\rho)+(d-1)^2$ as functions of $\rho$, for the three
coinciding Efimov points, $d$=$d_E$=$2.75$, where $|a_d|=\infty$.
The functions are asymptotically constant, approaching the values
$\xi_{d_E}^2 \approx -0.858$, $-2.679$, and $-0.270 $, 
for $m_H/m_L=1/1$, $133/6$, $87/39$, respectively. These negative 
constants are criteria for occurrence of the Efimov effect in all the three cases.

 The
dashed and dot-dashed curves are the same functions for the
neighboring $d$-values, $d=2.755$ and $d=2.745$, respectively.  
When $d>d_{E}$, as for $d=2.755$, the $\lambda^{(d)}$ functions
reproduce asymptotically the hyperspherical spectrum $K(K+2d-2)$
\cite{nie01}, which means that $\xi_d^2$ becomes positive for
sufficiently large $\rho$.  When $d<d_{E}$, as for $d=2.745$, 
the $\lambda^{(d)}$ functions diverge parabolically to $-\infty$ \cite{nie01},
and the large negative asymptotic values prevent  the appearance of 
bound states of large radii. Thus, there is no
Efimov-like states for $d$ outside a very narrow interval around
$d_E$.

\begin{figure}[t]
\centering
\includegraphics[width=0.85\linewidth]{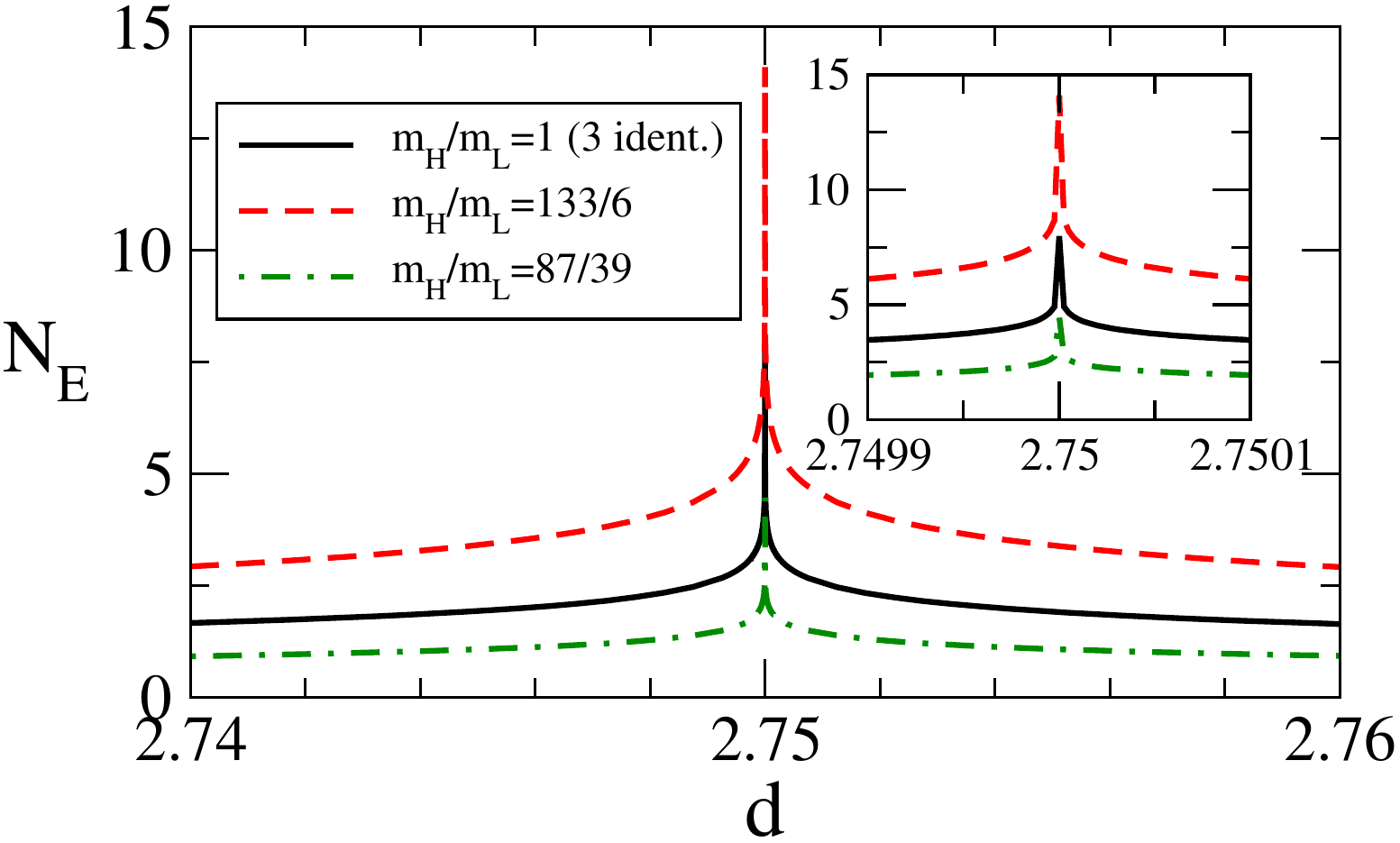}
\caption{For  $m_H/m_L=1/1$ (solid), $133/6$ (dashed), $87/39$
(dot-dashed), estimated number, $N_E$, of bound excited
states according to Eq.(\ref{num}) with $\rho_0 = b$ as functions of
$d$.  The narrow peaks, which occur for the Efimov point of
$d_{E}=2.75$, are zoomed in the inset.}
\label{fig3a}
\end{figure}

\section{Number of Efimov states}

As seen in Fig.~\ref{fig2}, the variation with $d$ of the effective
potentials in Eq.(\ref{rads1}) is very fast.  The consequence is that 
the Efimov effect, i.e., the Efimov states, appear and disappear
equally fast. We illustrate this in Fig.~\ref{fig3a}, where we show,
as a function of $d$, the estimated number, $N_E$, of bound states
across the Efimov point,  as given in Eq.(\ref{num}).
We notice the extremely fast increase and decrease of the number of bound states, as
highlighted in the inset.  The tiny dimension interval, $2.7499 < d <
2.7501$, i.e., a tiny variation in the confinement,
is enough to see essentially all the Efimov states appearing
and disappearing. It is important to keep in mind that, as shown in 
Fig.~\ref{fig2}, the strict Efimov conditions, i.e., asymptotic constant
effective potentials and $\xi_{d_E}^2<0$, are only fulfilled for $d=d_E$.
 We also emphasize that these features are precisely as extreme as observed
in the established Efimov scenario for $d=3$.

\begin{figure}[t]
\centering
\includegraphics[width=0.85\linewidth]{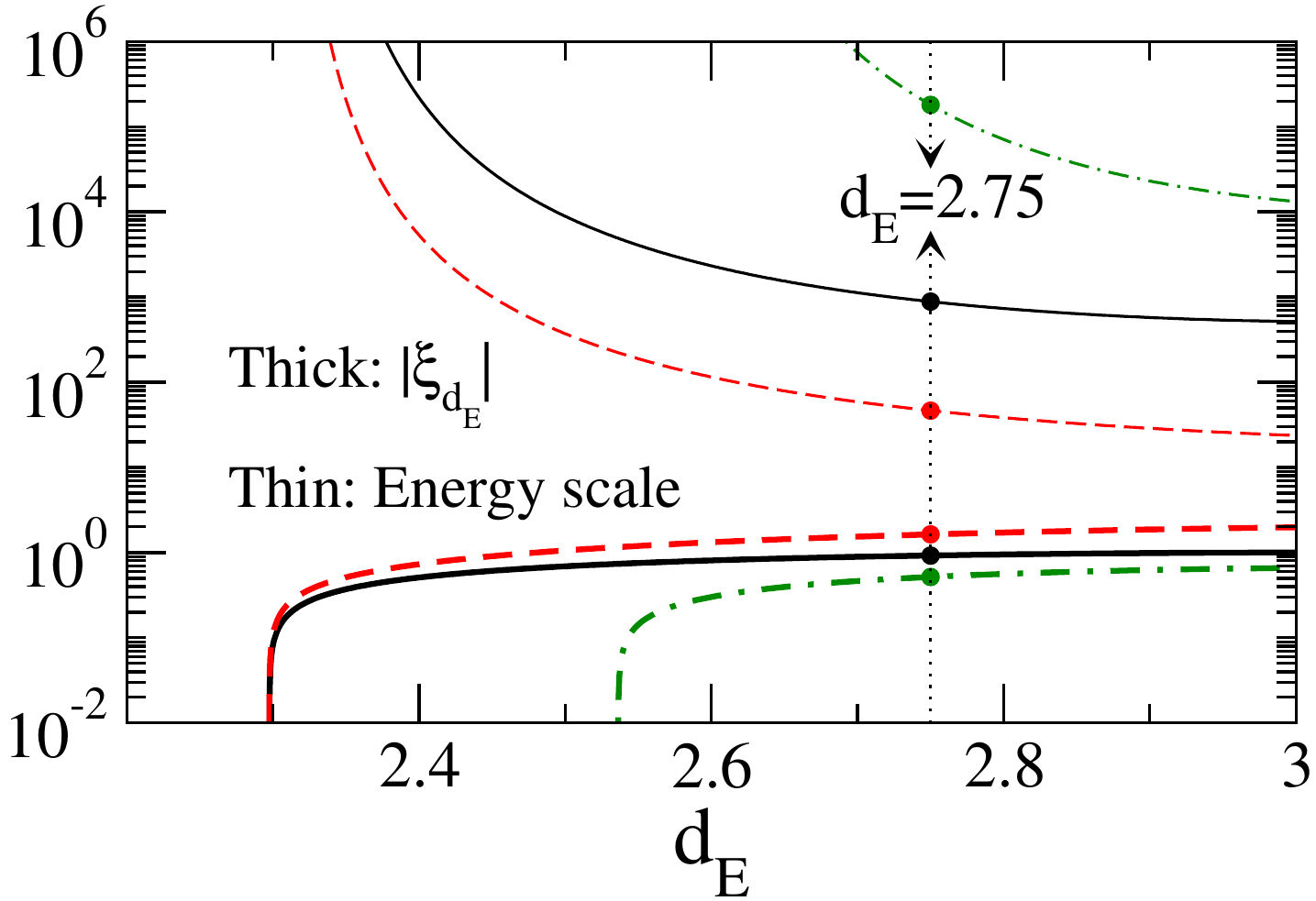}
\caption{For  $m_H/m_L=1/1$ (solid), $133/6$ (dashed), $87/39$
(dot-dashed),  value of $|\xi_{d_E}|$ as
a function of $d_E$ (thick) and the energy scale factor (thin) as
given in Eq.(\ref{scal}).}
\label{fig3b}
\end{figure}

In our calculations all the three Efimov points have arbitrarily been
chosen to be $d=d_E=2.75$.  However, the properties of the Efimov
states depend on the value of $d_E$.  Fortunately, the
large-distance constants, $\xi_{d_E}^2$, can be found
independent of the interactions through a transcendental equation as
explained in subsection 5.2.3 of Ref. \cite{nie01}.  As seen in
Fig.~\ref{fig3b} (thick curves), decreasing $d_E$ produces decreasing $|\xi_{d_E}|$,
which eventually becomes 0 for some $d_E=d_L$ limit value.  The Efimov effect
cannot occur for $d\leq d_L$, which for three identical bosons is
$d_L=2.298$, as already noted in \cite{nie01}.  For asymmetric
systems, $d_L$ oscillates depending on the mass ratio \cite{chr18}.
In particular, $d_L\approx 2.1$ for two non-interacting heavy and one
light particles if $m_H/m_L \gg 1$, while $d_L$ increases when
$m_H/m_L$ decreases.  For $m_H/m_L = 133/6$ we get $d_L\approx 2.3$,
as in the case of three identical particles, and $d_L\approx 2.54$ for
$m_H/m_L = 87/39$.

In Fig.~\ref{fig3b} we also show (thin curves) the energy scale factor,
Eq.(\ref{scal}), which for given $d_E$ decreases with increasing
$m_H/m_L$ ratio in the mass asymmetric case.  This crucial fact allows easier numerical
calculations of the Efimov states and facilitates as well experimental
detection.  This is evident in our case, $d_E=2.75$, where the energy
scale factors for $m_H/m_L=133/6$ and $87/39$ are 46.5 and 180214,
respectively, whereas we get 882.8 for the symmetric case of three identical particles
($m_L/m_H=1$).  This explains why the value of $N_E$ in
Fig.~\ref{fig3a} is always above and below the other two curves for
$m_H/m_L=133/6$ and $m_H/m_L=87/39$, respectively.

From the figure we can then conclude that, in the asymmetric case,  the smaller the
$m_H/m_L$ ratio the closer $|\xi_{d_E}|$ to zero and, therefore, the larger the energy
scale factor between the Efimov states. A reduction of the $m_H/m_L$ ratio makes soon
the Efimov states essentially unreachable both, theoretically and experimentally. As
an example, for the asymmetric case with $m_H/m_L=1$ the energy scales with a
factor of about $2.5\cdot 10^{10}$.

\section{Efimov energies}

\begin{figure}[t]
\centering
\includegraphics[width=0.9\linewidth]{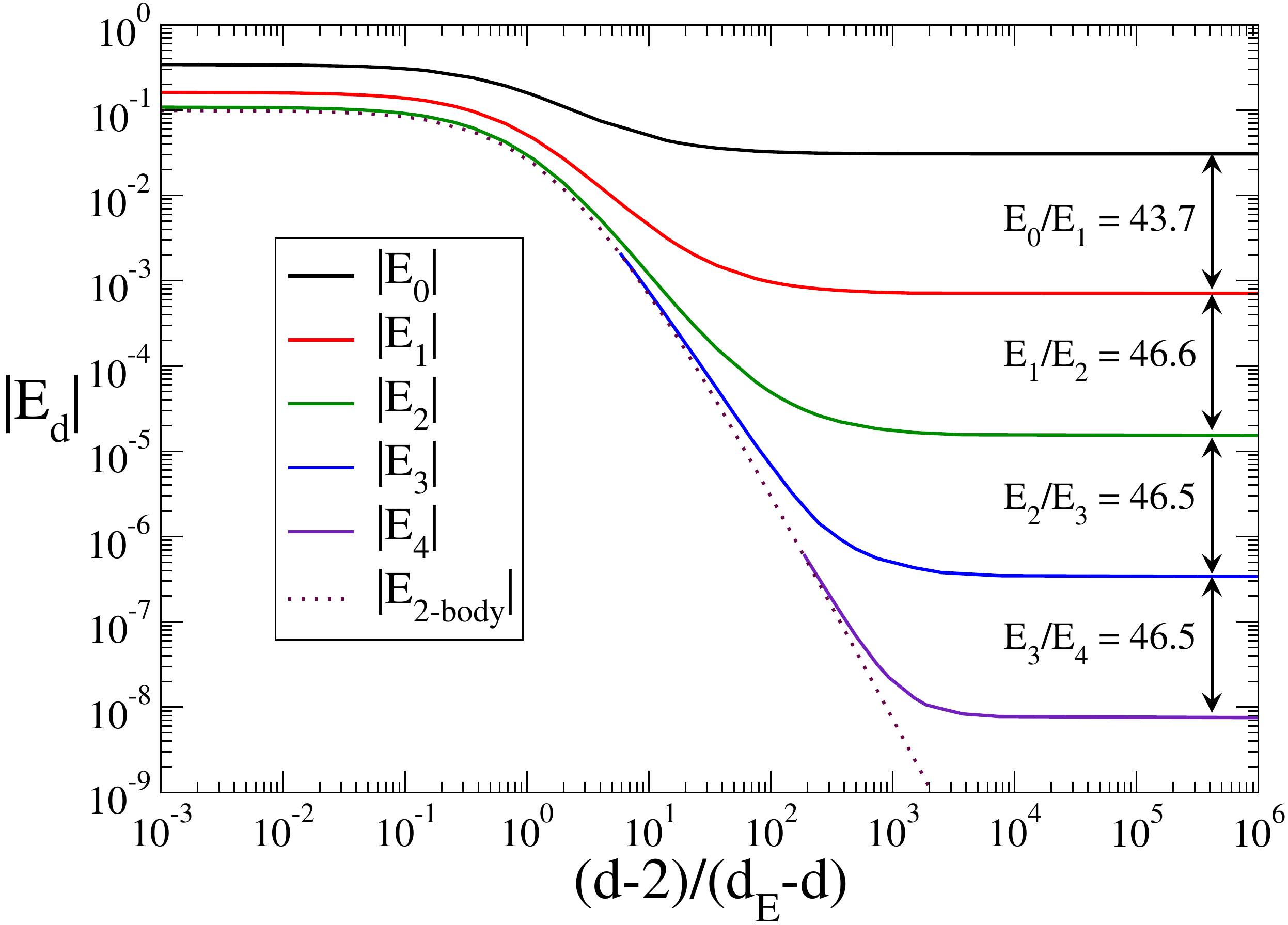}
\caption{Absolute values of the bound three-body energies, $|E_d|$, in units of
$\hbar^2/(\mu b^2)$, for
  the system of mass ratio, $133/6$, as functions of $(d-2)/(d_E-d)$,
  where $d_E=2.75$ is the Efimov point for this interaction.  The
  dotted curve correspond to the two-body binding energy.}
\label{fig4}
\end{figure}

The energies of the Efimov states appearing in the tiny $d$-interval
above  $d_E$ towards the Efimov point are shown in Fig.~\ref{fig4} for
the mass ratio $m_H/m_L=133/6$.  We show the lower side of the Efimov
point, $d<d_E$, where the two identical two-body subsystems are always bound.  
In order to expand the small interval for visibility, we show the
energies as functions of $(d-2)/(d_E-d)$.  For comparison we also show
the two-body binding energies in the figure (dotted curve).

The third and fourth excited states disappear along the two-body
threshold, whereas the ground state and the two first excited states
evolve as already shown by the dashed curves in Fig.~\ref{fig1}.  The
energies of the five computed Efimov states for $d=d_E$ are,
in units of $\hbar^2/(\mu b^2)$, $-3.050\cdot 10^{-2}$, $-6.985\cdot 10^{-4}$,
$-1.498\cdot10^{-5}$, $ -3.225\cdot 10^{-7}$, and
$-6.931\cdot10^{-9}$, respectively.  The energy scale factors are then equal to
43.7, 46.6, 46.5, and 46.5, in very good agreement with the value
($\approx 46.47$) obtained from Eq.(\ref{scal}) for
$\xi_{d_E}^2=-2.679$ (see Fig.\ref{fig2}).

In the higher side of the Efimov point, $d>d_E$, where the two identical 
two-body subsystems are unbound, the Efimov states disappear
in the continuum, as shown in Fig.~\ref{fig1} for the two lowest states. 
Beyond $d_E$ the energy of the different states drops rapidly, and eventually,
for the interaction chosen in this work, the three-body states become unbound.

\section{Conclusions}

In this work we have explained a novel procedure that permits to obtain the 
Efimov conditions for a given three-body system.
The tool is an external, directional squeezing, one-body potential.
The usual external confining field, always present in cold atom
experiments, must be deformed for our purpose, where a sufficiently
strong deformation would squeeze the system continuously into two
dimensions.  The particles must then eventually adjust to a two-dimensional world,
where even a weak two-body attraction supports a bound state.

Thus, for an interaction unable to bind the two-body system in three dimensions, 
at some intermediate 
deformation the two-body system must change
from unbound to being bound.  This Efimov point, with zero two-body
binding energy, corresponds to an infinite scattering length, which 
leads to the infinite series of three-body Efimov states.  As in
three dimensions, this structure will emerge for any ground, or
excited-state vanishing, two-body energy.  The characteristic features
are precisely the same as in three dimensions, and therefore open to
the same type of tests.

The advantage now is that two independent controllable parameters 
are then available for tuning to the Efimov condition, that is
Feshbach tuning and external field squeezing.  This must in any case
be more flexible than exploiting only one of these degree-of-freedom.
It could be that an additional two-body subsystem can be made to
contribute, perhaps the Feshbach tuning can be less precise, perhaps
the Efimov scaling can be optimized to a denser spectrum with more
opportunities, perhaps different systems can be studied, or perhaps
applications on different problems present advantages.

We employed the recently formulated $d$-method, which is precisely as
easy, or difficult, as an ordinary three-body calculation.  It is
equivalent to the brute force method of applying an external field,
where one more three dimensional degree-of-freedom is required in more
complicated calculations.  The present results are obtained for
identical bosons and for distinguishable particles.  Additional
applications are abundant, as for example more particles, different
quantum symmetries, squeezing more than one dimension, or asymmetric squeeze 
of the dimensions.

\section*{Acknowledgments} This work has been partially supported by 
the Spanish Ministry of Science, Innovation and University 
MCIU/AEI/FEDER,UE (Spain) under Contract No. PGC2018-093636-B-I00.

\end{document}